\documentclass[12pt]{article}
\usepackage{graphicx}
\usepackage{amsfonts}
\usepackage{amssymb}
\usepackage{mathrsfs}
\usepackage{amsmath}
\usepackage[unicode]{hyperref}
\usepackage[numbers,sort&compress]{natbib}

\begin{document}
\renewcommand{\thefootnote}{\fnsymbol{footnote}}
\begin{titlepage}

\vspace{10mm}
\begin{center}
{\Large\bf Thermodynamics of the Schwarzschild-AdS black hole with a minimal length}
\vspace{8mm}

{{\large Yan-Gang Miao${}^{1,2,}$\footnote{E-mail address: miaoyg@nankai.edu.cn} and Yu-Mei Wu${}^{1,}$\footnote{E-mail address: wuym@mail.nankai.edu.cn}}\\

\vspace{6mm}
${}^{1}${\normalsize \em School of Physics, Nankai University, Tianjin 300071, China}

\vspace{3mm}
${}^{2}${\normalsize \em Max-Planck-Institut f\"ur Gravitationsphysik (Albert-Einstein-Institut),\\
M\"uhlenberg 1, D-14476 Potsdam, Germany}

}

\end{center}

\vspace{10mm}
\centerline{{\bf{Abstract}}}
\vspace{6mm}
%\noindent
Using the mass-smeared scheme of black holes, we study the thermodynamics of black holes. Two interesting models are considered. One is the self-regular Schwarzschild-AdS black hole whose mass density is given by the analogue to probability densities of quantum hydrogen atoms. The other model is the same black hole but whose mass density is chosen to be a rational fractional function of radial coordinates. Both mass densities are in fact analytic expressions of the ${\delta}$-function. We analyze the phase structures of the two models by investigating the heat capacity at constant pressure and the Gibbs free energy in an isothermal-isobaric ensemble. Both models fail to decay into the pure thermal radiation even with the positive Gibbs free energy due to the existence of a minimal length. Furthermore, we extend our analysis to a general mass-smeared form that is also associated with the ${\delta}$-function, and indicate the similar thermodynamic properties for various possible mass-smeared forms based on the ${\delta}$-function.

\vskip 20pt
\noindent
{\bf PACS Number(s)}: 04.60.Bc, 04.70.Dy

\vskip 10pt
\noindent
{\bf Keywords}: Self-regular Schwarzschild-AdS black hole, general mass density based on the ${\delta}$-function, thermodynamics

\end{titlepage}
\newpage
\renewcommand{\thefootnote}{\arabic{footnote}}
\setcounter{footnote}{0}
\setcounter{page}{2}

\section{Introduction}

Non-renormalizability is a well-known puzzle when gravity is combined with quantum theory. Among many attempts to solve the problem, the idea of a non-perturbative quantum gravity theory is attractive. Recently, Dvali and collaborators propelled~\cite{1005.3497,1010.1415} this idea by putting forward the ``UV self-complete quantum gravity" in which the production of micro black holes is assumed to play a leading role in scattering at the Planck energy scale. According to the Heisenberg uncertainty relation, an incident photon needs to have higher energy in order to locate a target particle more accurately. When the photon energy is comparable to that of the particle, it is likely to create a new particle. If the energy of the incident photon gets further higher, say the Planck energy, so huge energy confined in a small scale may create a black hole, the so-called the hoop conjecture~\cite{1311.5698}. After the formation of micro black holes, the higher the incident photon energy is, the bigger the black hole is, which makes the probe meaningless. As a result, the existence of a minimum length may be deduced by the horizon radius of an extreme black hole.

Regular black holes, i.e. the black holes without singularity at the origin, whose research could be traced back to the Bardeen's brilliant work~\cite{bardeen}, appeared in refs.~\cite{0310151,0506126} on the discussions of mass definition, casual structure, and other related topics. In 2005, for the purpose of constructing the noncommutative geometry inspired Schwarzschild black hole, Nicolini, Smailagic, and Spallucci suggested~\cite{0510112} that the energy-momentum tensor should be modified in the right hand side of Einstein's equations, such that it describes a kind of anisotropic fluid rather than the perfect fluid, where the original point-like source depicted by the $\delta$-function should be replaced by a mass-smeared distribution, while no changes should be made in the left hand side. By replacing the $\delta$-function form of the point-like source by the Gaussian form\footnote{Here the Gaussian form can be regarded as an analytic expression of the ${\delta}$-function. In general, an analytic expression of the ${\delta}$-function is such an elementary function that approaches the ${\delta}$-function under a limit of a parameter. For instance, the mass density of our first model, see eq.~(\ref{eq:density1}), is the ${\delta}$-function under the limit $a \to 0$, that is, $\lim_{a \to 0}\frac{ e^{-\frac{r}{a}}}{8 \pi  a^3}=\delta^{(3)}(r)$, where $\delta^{(3)}(r)$ is defined as $\int_{0}^{\infty} \delta^{(3)}(r)\,4\pi r^2 dr=1$.} and then solving the modified Einstein equations, they obtained a self-regular Schwarzschild metric and established a new relation between the mass and the horizon radius, i.e. the existence of a minimum radius and its corresponding minimum mass. They also discussed the relevant thermodynamic properties and deduced the vanishing temperature in the extreme configuration, and thus eliminated the unfavorable divergency of the Hawking radiation. Besides, there is a correction to the entropy in the near-extreme configuration. Since then, a lot of related researches have been carried out, such as those extending to high dimensions~\cite{1202.1686,1511.00853}, introducing the AdS background~\cite{1212.5044}, quantizing the mass of black holes~\cite{1503.01681}, and generalizing to other types of black holes~\cite{0612035,1003.3918}, etc.

In this paper, we at first introduce the AdS spacetime background to the self-regular Schwarzschild black hole and then investigate equations of state and phase transitions of the self-regular Schwarzschild-AdS black hole with two specific mass densities, where one mass density is assumed~\cite{1511.04865} by an analogue to a quantum hydrogen atom and the other to a collapsed shell.
%Very recently, we suggested a new scheme to quantize the mass of black holes by an analogue to hydrogen atoms, in which we related the mass density of black holes with the wave function of hydrogen atoms. Despite of the quantization, this analogy offers us an approach to constructing regular black holes in the semi-classical framework, i.e. the density distribution taking an exponential form instead of the $\delta$-function.
Although the mass densities of the two models are different, both of them can be regarded as a smeared point-like particle and then be depicted by an analytic expression of the ${\delta}$-function. Next,
we extend our discussions to a general mass-smeared form which can be written as an unfixed analytic expression of the ${\delta}$-function, and explain the similarity in thermodynamic properties for different mass distributions previously studied in refs.~\cite{1212.5044, 1110.5332, 1410.1706}.

The paper is organized as follows. In section 2, we derive the relations between masses and horizons of the two models, and give the limits of vacuum pressure under the consideration of the hoop conjecture. Then,  we analyze the thermodynamic properties of the two models in detail in section 3.  We turn to our analysis in section 4 for the self-regular Schwarzschild-AdS black hole with a general mass distribution. Finally, we give a brief summary in section 5.

We adopt the Planck units in this article: $\hbar=c=G=k_B=1$.

\section{Two specific models}

\subsection{Mass density based on analogy between black holes and quantum hydrogen atoms}
\begin{figure}[!ht]
\centering\includegraphics[height=6cm, width=16cm]{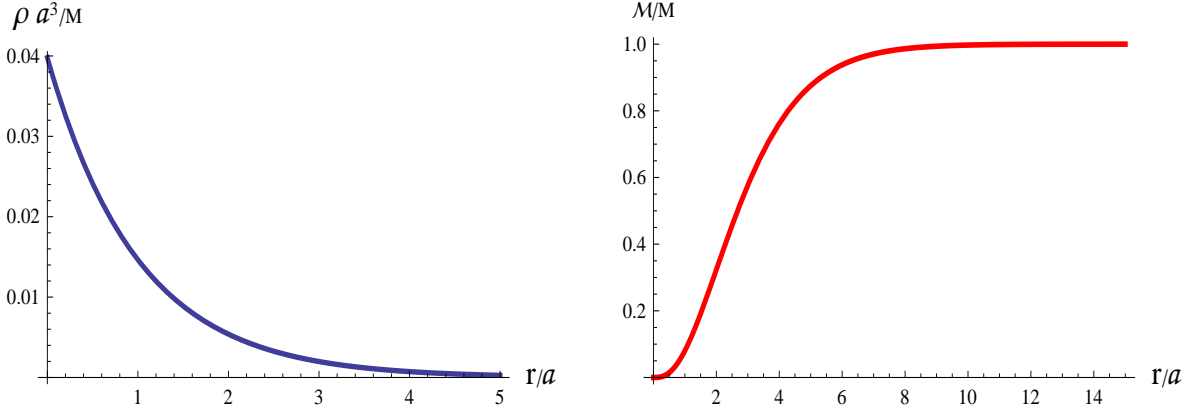}
\caption{Plots of the mass density eq.~(\ref{eq:density1}) (left) and of the mass distribution eq.~(\ref{eq:M_e}) (right). Note that, when we take the limit $a\rightarrow0$, except the origin with $\mathcal{M}(0)=0$, the mass ratio $\mathcal{M}(r)/M$ approaches 1 at other radii due to $r/a\rightarrow \infty$, which indicates that the mass distribution goes to the step function.}
\label{fig:rho_1}
\end{figure}

\begin{figure}[!ht]
\centering\includegraphics[height=7cm]{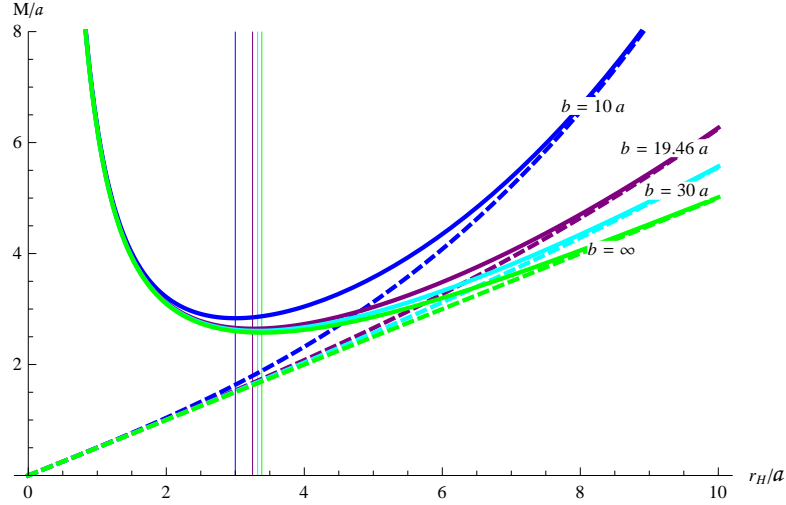}
\caption{Plot of the relation of $M$ with respect to $r_{\rm H}$. The solid curves correspond to the relation eq.~(\ref{eq:M_r_e}), the dashed curves correspond to the relation of the ordinary Schwarzschild-AdS black hole with $M=\frac{r_{\rm H}}{2}(1+\frac{r_{\rm H}^2}{b^2})$, and each vertical line lies at the extremal radius of the curve with the same color. The blue, purple, cyan, and green colors correspond to the cases of $b/a=10$, $19.46$ (the critical vacuum pressure), $30$, and $\infty$, respectively.}
\label{fig:M_r_e}
\end{figure}

Based on our recent work in which we made an analogy between black holes and quantum hydrogen atoms~\cite{1511.04865}, we choose the probability density of the ground state of hydrogen atoms as the mass density of black holes,
\begin{align}
\rho(r)=M\frac{ e^{-\frac{r}{a}}}{8 \pi  a^3},\label{eq:density1}
\end{align}
where $M$ is the total mass of black holes and $a$ with the length dimension will be seen to be related to a minimal length through the extreme configuration discussed below. We plot the mass density in Figure \ref{fig:rho_1} and obtain $\rho(0)=M/(8\pi a^3)$, which means that the origin is no longer singular. Now the mass distribution of black holes takes the form,
\begin{align}
\mathcal{M}(r)=\int_0^r\rho({\tilde r})\,4\pi {\tilde r}^2\,\mathrm{d}{\tilde r}=M\left[1-e^{-\frac{r}{a}}\left(1+\frac{r}{a}+\frac{r^2}{2a^2}\right)\right],
\label{eq:M_e}
\end{align}
which, divided by $M$, can be understood as a kind of step functions with continuity  (see the right diagram in Figure \ref{fig:rho_1}). It is easy to see that such a function $\mathcal{M}(r)/M$ goes to the step function under the limit $a \to 0$.
Correspondingly, as done in refs.~\cite{0510112,1212.5044}, we work out the self-regular Schwarzschild-AdS metric through solving the modified Einstein equations associated with the above mass distribution,
\begin{align}
\mathrm{d}s^2=-\left(1-\frac{2\mathcal{M}(r)}{r}+\frac{r^2}{b^2}\right)\mathrm{d}t^2+\left(1-\frac{2\mathcal{M}(r)}{r}+
\frac{r^2}{b^2}\right)^{-1}\mathrm{d}r^2+r^2\,\mathrm{d}\Omega^2,
\label{eq:ds^2_ads}
\end{align}
where $b$ represents the curvature radius of the AdS spacetime and has the relation with the the vacuum pressure as follows,
\begin{align}
P\equiv-\frac{\Lambda}{8\pi}=\frac{3}{8\pi b^2}.
\label{eq:P}
\end{align}
One can easily find that the self-regular metric goes back to the ordinary Schwarzschild-AdS one if $r\gg a$, and that on the other hand it goes to the de Sitter form if $r$ approaches zero,
\begin{align}
\mathrm{d}s^2=-\left(1-\frac{M }{3 a^3}r^2+\frac{r^2}{b^2}\right)\mathrm{d}t^2+\left(1-\frac{M }{3 a^3}r^2+\frac{r^2}{b^2}\right)^{-1}\mathrm{d}r^2+r^2 \,\mathrm{d}\Omega^2,\label{dSmetric}
\end{align}
where the equivalent cosmological constant can be understood as,\footnote{The quantity is inferred to be positive from the ranges of $M$ and $b$ in the following discussion, which will be mentioned at the end of subsection 3.1.} $\tilde\Lambda \equiv \frac{M}{a^3}-\frac{3}{b^2}$. Thus the singularity is avoided since the positive cosmological constant represents the negative (outward) vacuum pressure that prevents the black hole being less than a certain scale from collapsing.
From $g_{00}=0$, one can have the relation between the total mass $M$ and the horizon radius $r_{\rm H}$,
\begin{align}
M=\frac{r_{\rm H}}{2}\left(1+\frac{r_{\rm H}^2}{b^2}\right)\left[1-\left(1+\frac{r_{\rm H}}{a}+\frac{r_{\rm H}^2}{2a^2}\right)e^{-\frac{r_{\rm H}}{a}}\right]^{-1},
\label{eq:M_r_e}
\end{align}
which  is shown in Figure \ref{fig:M_r_e} under different vacuum pressure.

From Figure \ref{fig:M_r_e}, we observe that there is a lower bound $M_0$ for the black hole mass, which is regarded as the mass of the extreme black hole. For the case of $M>M_0$, the black hole possesses two horizons; for the case of $M=M_0$, the two horizons coalesce into one, $r_{\rm H}=r_0$, as the extremal radius, i.e. the minimal length since no black holes are smaller than the extreme black hole. The extremal radius depends on the vacuum pressure, which can be determined by $\frac{\partial M}{\partial r_{\rm H}}\big|_{r_{\rm H}=r_0}=0$, where
\begin{align}
\frac{\partial M}{\partial r_{\rm H}}=\frac{ae^{r_{\rm H}/a}[2a^3(e^{r_{\rm H}/a}-1)(b^2+3r_{\rm H}^2)-2a^2r_{\rm H}(b^2+3r_{\rm H}^2)-ar_{\rm H}^2(b^2+3r_{\rm H}^2)-r_{\rm H}^3(b^2+r_{\rm H}^2)]}{b^2[-2a^2(e^{r_{\rm H}/a}-1)+2ar_{\rm H}+r_{\rm H}^2]^2}.
\end{align}
It is difficult to express $r_0$ as the function of $b$ since the equation is transcendental, so we solve $b$ with respect to $r_0$ instead,
\begin{align}
b=r_0\sqrt{-3+\frac{2r_0^3/a^3}{2+2r_0/a+r_0^2/a^2+r_0^3/a^3-2e^{r_0/a}}}.
\label{eq:b_e}
\end{align}

Considering the hoop conjecture~\cite{1311.5698} where the mean radius ${\bar r}$ is supposed to be not greater than the extremal radius $r_0$, we have estimated the ranges of both $b$ and $r_0$ in ref.~\cite{1511.04865} through the numerical fitting. It is easy to calculate
\begin{align}
{\bar r}=\frac{1}{M}\int_0^\infty r \rho(r) \,4\pi r^2\,\mathrm{d}r=3 a.
\label{eq:hoop_e}
\end{align}
The hoop conjecture implies ${\bar r} \leq r_0$, which ensures the formation of a black hole. Using eq.~(\ref{eq:b_e}) and the hoop conjecture, one can obtain the
exact ranges of the ratios, $b/a$: [9.996, $\infty$), $r_0/a$: [3, 3.384), and $M/a$: (2.575, 2.835], respectively. We note that a smaller $b$ corresponds to an extreme black hole with a smaller radius but a larger mass, i.e. when $b/a=9.996$, we have $r_0/a=3$ and $M/a=2.835$. % (intuitive from Figure \ref{fig:M_r_e}).
The reason is that a smaller $b$ represents the larger vacuum pressure which makes the matter denser~\cite{1511.04865}.

\subsection{Mass density based on a collapsed shell}
\begin{figure}[!ht]
\centering\includegraphics[height=6cm,width=16cm]{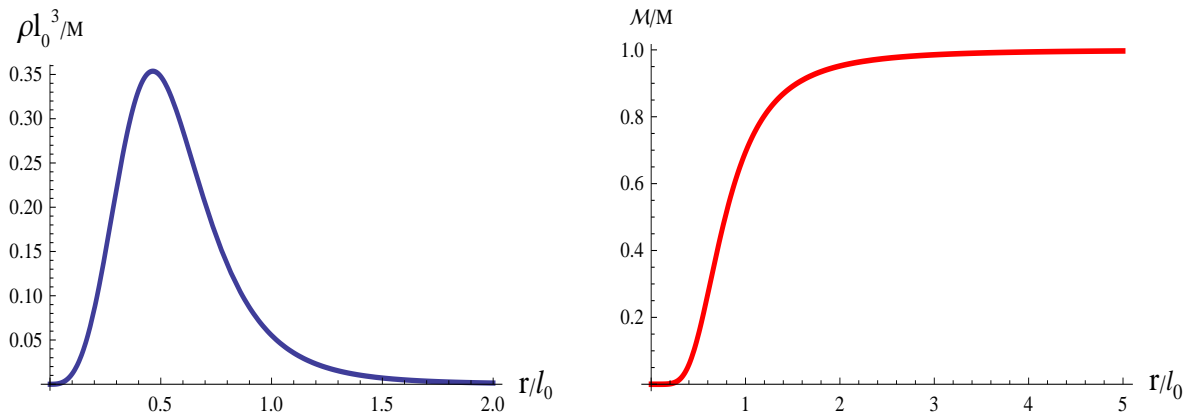}
\caption{Plots of the mass density eq.~(\ref{eq:density2}) (left) and of the mass distribution eq.~(\ref{eq:M_s}) of the collapsed shell.}
\label{rho_2}
\end{figure}

Besides collapsed cores, like the first model, a collapsed shell is also an important model of black hole formation, which is usually dealt with as a massive membrane without thickness. However, if the minimal length is considered, the shell is supposed to be described with a smeared distribution~\cite{1110.5332}. One can take the following rational fractional function of radial coordinators as the mass density of black holes,
\begin{align}
\rho(r)=\frac{75 M l_0^3 r^3}{2 \pi  \left(l_0^3+5r^3\right)^3},
\label{eq:density2}
\end{align}
where $l_0$ is a quantity with the length dimension which can be regarded as the minimal length, and $M$ the total mass of back holes. The relation is plotted in Figure \ref{rho_2}.

\begin{figure}[!ht]
\centering\includegraphics[height=7cm]{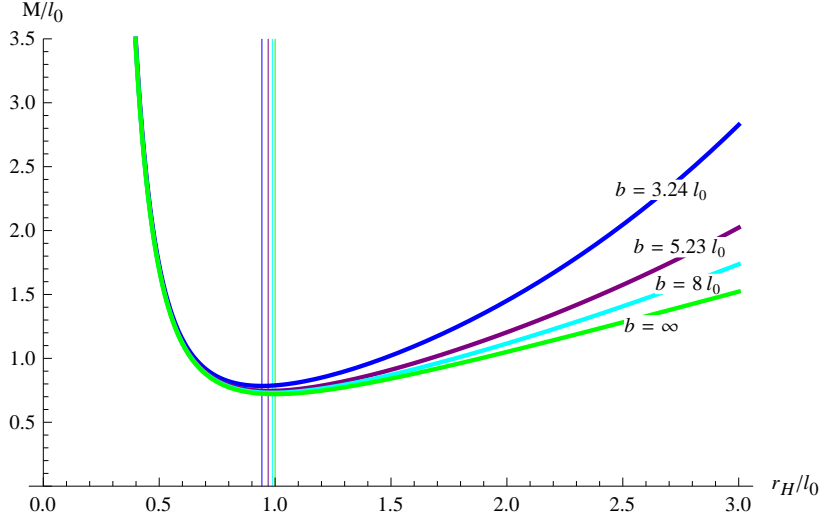}
\caption{Plot of the relation of $M$ with respect to $r_{\rm H}$. Each vertical line lies at the extremal radius of the curve with the same color.}
\label{M_s}
\end{figure}

The mass distribution of this model, $\mathcal{M}(r)$, thus has the form,
\begin{align}
\mathcal{M}(r)=\int_0^r\rho({\tilde r})\,4\pi {\tilde r}^2\,\mathrm{d}{\tilde r}=\frac{25 M r^6}{\left(l_0^3+5 r^3\right)^2},
\label{eq:M_s}
\end{align}
which, like eq.~(\ref{eq:M_e}), can also be understood as a kind of step functions with continuity and goes to the step function under the limit $l_0 \to 0$.
By solving $g_{00}=0$, we get the total mass $M$ as the function of the horizon radius $r_{\rm H}$,
\begin{align}
M=\frac{\left(b^2+r_{\rm H}^2\right) \left(l_0^3+5 r_{\rm H}^3\right)^2}{50 b^2 r_{\rm H}^5},
\label{eq:M}
\end{align}
which is plotted in Figure \ref{M_s}.
As done for the first model, one can also determine the allowed ranges of the ratios  for the extreme configuration in this model, $b/l_0$: [3.24, $\infty$), $r_0/l_0$: [0.943, 1), and $M/l_0$: (0.720, 0.784], respectively, when the hoop conjecture is considered.
Moreover, the data lead to the same result as in the first model, i.e. a smaller extreme black hole is heavier than a greater one.

\section{Thermodynamics}
\subsection{Thermodynamics of hydrogen-atom-like black holes  }
In this subsection we are going to analyze the thermodynamic properties of the hydrogen-atom-like black holes, especially new phase transitions associated with the minimal length. %At first we get a glimpse of them from the equation of state.
\subsubsection{Equation of state and entropy}
\begin{figure}[!ht]
\centering\includegraphics[height=7cm]{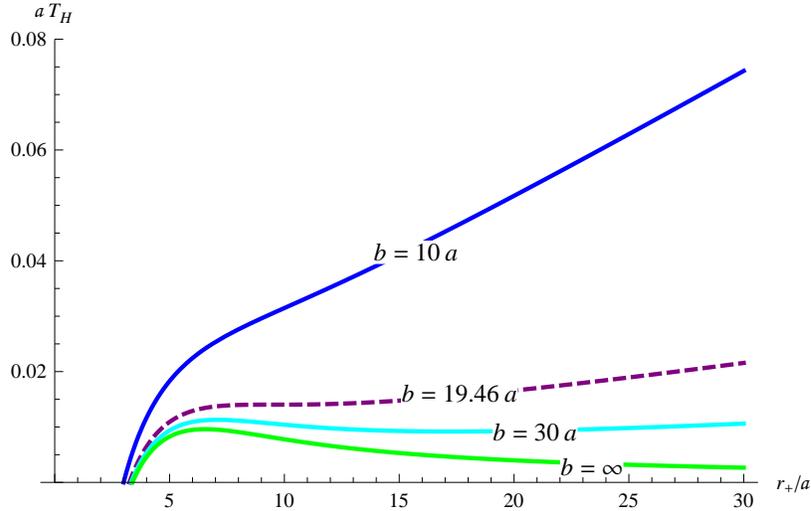}
\caption{Plot of the Hawking temperature $T_{\rm H}$ with respect to $r_+$ under different vacuum pressure. The blue, cyan, and green solid curves correspond to the cases of $b/a=10$, $30$, and $\infty$, respectively, and the extremal radii in the three cases take $r_0/a=3.000$, $3.325$, and $3.384$, respectively. The dashed purple curve corresponds to $b/a=19.46$ and $r_0/a=3.253$, i.e. the critical case where the maximum and minimum of temperature coalesce.}
\label{fig:T_H_e}
\end{figure}

We start the discussion of the thermodynamics by calculating the Hawking temperature:
\begin{align}
T_{\rm H}\equiv \frac{\kappa_{\rm H}}{2\pi}=\frac{1}{4\pi \sqrt{-g_{00}g_{11}}}\left|\frac {\mathrm{d}g_{00}}{\mathrm{d}r}\right|_{r=r_+},
\label{eq:T_H(kappa)}
\end{align}
where $\kappa_{\rm H}$ is the surface gravity and $r_+$ the outer horizon radius.
Using eqs.~(\ref{eq:ds^2_ads}) and (\ref{eq:M_e}), we obtain
\begin{align}
T_{\rm H}=\frac{1}{4 \pi }\left[\frac{1}{r_+}+\frac{3r_+}{b^2}+\frac{r_+^2}{2a^3}\left(1+\frac{r_+^2}{b^2}\right)\left(1+\frac{r_+}{a}+\frac{r_+^2}{2a^2}-e^{r_+/a}\right)^{-1}\right],
\label{eq:T_H_e}
\end{align}
which is plotted in Figure \ref{fig:T_H_e}.
 These curves show one common property that the temperature of the extreme black hole vanishes, which can be strictly proved by substituting eq.~(\ref{eq:b_e}) into eq.~(\ref{eq:T_H_e}). This property means that the evaporation of black holes behaves well, i.e.  it has no divergency due to the appearance of the minimal length~\cite{0510112}. In addition, the curves display different characteristics when the vacuum pressure increases,  which implies various phase transitions that will be discussed in the next subsection. The phase transitions can be classified into four types: $(\mathrm{i})$ For the zero pressure case~($b \to \infty$, see, for instance, the green curve of Figure \ref{fig:T_H_e}), there exists one maximum of temperature, and the temperature approaches zero  when $r_+\to \infty$; $(\mathrm{ii})$ For the relatively low pressure~($19.46a<b<\infty$, see, for instance, the cyan curve of Figure \ref{fig:T_H_e}), the temperature has one additional minimum following one maximum, and then it increases when $r_+$ increases; $(\mathrm{iii})$ For the critical pressure $P_c$~($b=19.46a$, see, for instance, the dashed purple curve of Figure \ref{fig:T_H_e}), the maximum and minimum merge into one inflexion; $(\mathrm{iv})$ For the relatively high pressure~($9.996a<b<19.46a$, see, for instance, the blue curve of Figure \ref{fig:T_H_e}), the temperature rises smoothly as the horizon radius increases.

\begin{figure}[!ht]
\centering\includegraphics[height=7cm]{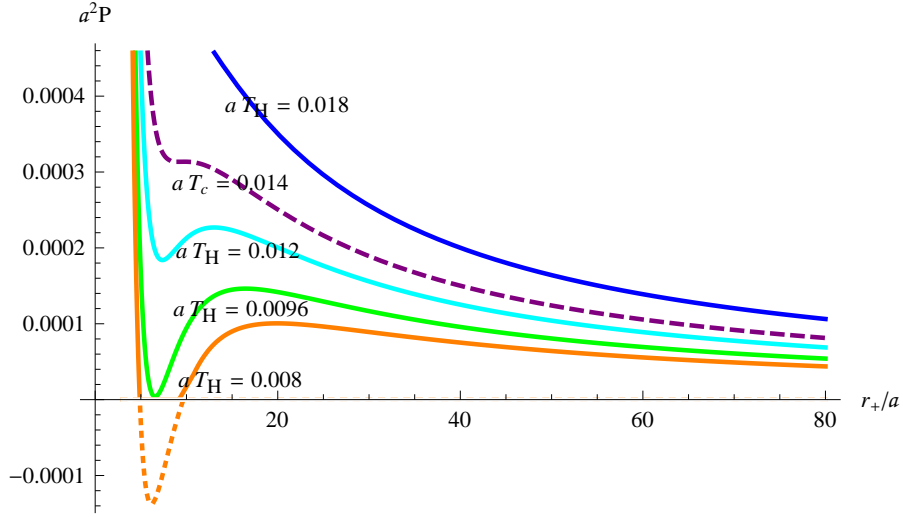}
\caption{Plot of the equation of state. The solid curves correspond to the cases of $aT_{\rm H}=0.018$ (blue), $0.012$ (cyan), and $0.0096$ (green) from top to bottom, respectively. The orange curve corresponds to $aT_{\rm H}=0.008$, but the dotted part, where the pressure is negative, does not actually exist. The dashed purple curve corresponds to the critical temperature $aT_c=0.014$.}
\label{P_v}
\end{figure}

In fact, eq.~(\ref{eq:T_H_e}) is the equation of state and Figure \ref{fig:T_H_e} is the diagram of isobar in the temperature-volume plane.\footnote{The radius and the specific volume have a simple relation, $v=2 l_p^2 r_+$, where $l_p$ is the Planck length, suggested in ref.~\cite{1205.0559} by both the dimensional analysis and the analogy between the RN-AdS black hole and the van der Waals fluid.\label{2}} One can rewrite eq.~(\ref{eq:T_H_e}) in a familiar way by using eq.~(\ref{eq:P}),
\begin{align}
P=-\frac{1}{8\pi r_+^2}+\frac{T_{\rm H}}{2r_++\frac{r_+^4}{3a^3}\left(1-e^{\frac{r_+}{a}}+\frac{r_+}{a}+\frac{r_+^2}{2a^2}\right)^{-1}}-\frac{r_+}{4\pi\left[6a^3\left(1-e^{\frac{r_+}{a}}+\frac{r_+}{a}+\frac{r_+^2}{2a^2}\right)+r_+^3\right]},
\label{eq:p_V}
\end{align}
and plot eq.~(\ref{eq:p_V}) as the diagram of isotherm in the pressure-volume plane shown in Figure \ref{P_v}. The diagram reveals the similarity to that of the van der Waals fluid, implying the occurrence of phase transitions in the range from zero temperature to the critical temperature $T_c=0.014/a$.

A convincing demonstration was given in ref.~\cite{0904.2765}, where the black hole mass $M$ is regarded as the enthalpy.  Alternatively, an intuitive explanation was provided in ref.~\cite{1008.5023}. That is, the negative cosmological constant corresponds to the positive pressure and thus to the negative vacuum energy density $\epsilon$, where $\epsilon +P=0$. Therefore,  the total energy is supposed to include the vacuum energy, i.e. $E=M+\epsilon V=M-PV$, and hence $M$ is the enthalpy $H$ rather than the internal energy of black holes. As a result, the first law associated with the  AdS spacetime is modified to be
\begin{align}
\mathrm{d}M=T_{\rm H}\mathrm{d}S+V\mathrm{d}P.
\label{eq:dM}
\end{align}

\begin{figure}[!ht]
\centering\includegraphics[height=7cm,width=16cm]{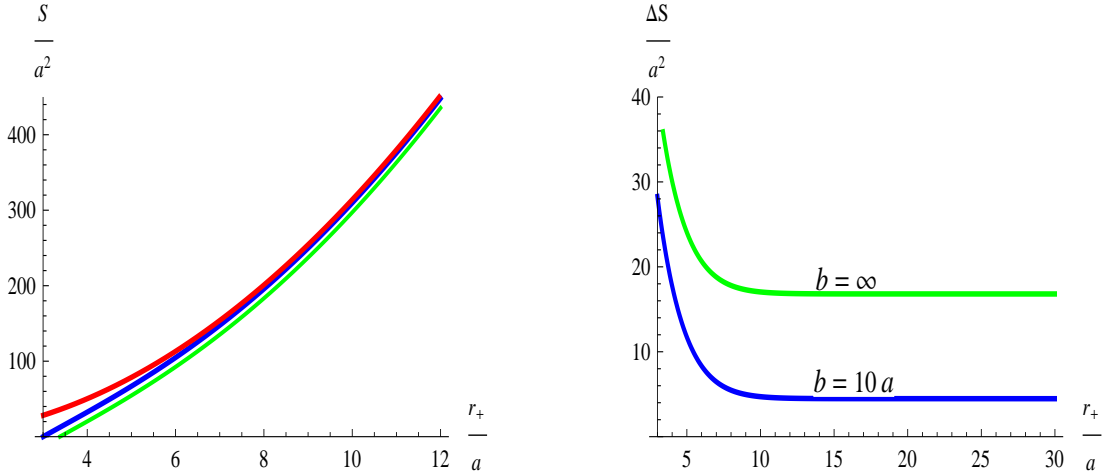}
\caption{Plots of the entropy $S$ (left) and of $\Delta S \equiv S-\pi r_+^2$ (right) with respect to $r_+$. The red curve in the left diagram corresponds to the entropy $\pi r_+^2$ of the ordinary Schwarzschild-AdS situation; the blue and green curves in both diagrams correspond to the cases of $b/a=10$ and $b=\infty$, respectively.}
\label{S_e}
\end{figure}

It is evident that the second term in eq.~(\ref{eq:dM}) vanishes for the fixed cosmological constant or constant pressure. Thus we can derive the entropy from eqs.~(\ref{eq:M_r_e}) and (\ref{eq:T_H_e}),
\begin{align}
S=\int^{r_+}_{r_0}\frac{\mathrm{d}M}{T_{\rm H}}=\int^{r_+}_{r_0}\frac{2\pi r}{1-e^{-\frac{r}{a}}\left(1+\frac{r}{a}+\frac{r^2}{2a^2}\right)}\mathrm{d}r,
\label{eq:S}
\end{align}
where the integration is performed from $r_0$ in order to guarantee the vanishing entropy at zero temperature~\cite{1212.5044}. It is hard to integrate eq.~(\ref{eq:S}) analytically but much easier to plot it. As shown in Figure \ref{S_e}, for the near-extreme configuration, the entropy has an obvious deviation from the area law, but with the horizon radius increasing, the deviation is getting smaller and smaller and finally approaches a constant. Note that such a constant is varying for different pressure or for different $b$ parameters because the lower bound of the integration $r_0$, i.e. the horizon radius of extreme black holes depends on the $b$ parameter, see the right diagram of Figure \ref{S_e}.

\subsubsection{Phase transition}
\begin{figure}[!ht]
\centering\includegraphics[height=10cm]{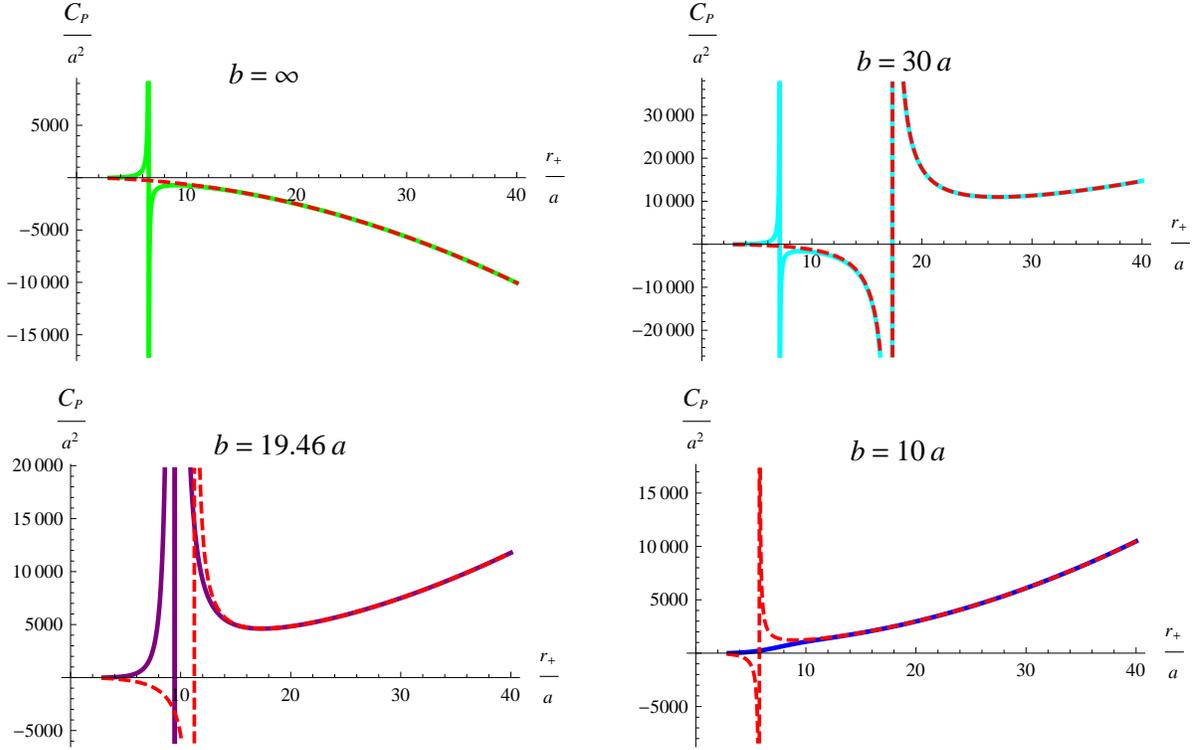}
\caption{Plots of the heat capacity $C_P$ with respect to $r_+$ under different pressure. They correspond to $b/a \to\infty$~(top left), $b/a=30$~(top right), $b/a=19.46$~(the critical vacuum pressure, bottom left), and $b/a=10$~(bottom right), respectively. The solid curves correspond to eq.~(\ref{eq:C_P}), and the dashed red curves correspond to the relationship $C_P=\frac{2\pi r_+^2(b^2+3r_+^2)}{3r_+^2-b^2}$ of the ordinary  Schwarzschild-AdS situation.}
\label{C_P}
\end{figure}

We discuss the phase transition of the hydrogen-atom-like model in this subsection. The heat capacity at constant pressure takes the form,
\begin{align}
C_P\equiv\left(\frac{\partial H}{\partial T_{\rm H}}\right)_P=\left(\frac{\partial M}{\partial T_{\rm H}}\right)_P=\left(\frac{\partial M}{\partial {r_+}}\right)\left(\frac{\partial T_{\rm H}}{\partial {r_+}}\right)^{-1}.
\label{eq:C_P}
\end{align}
It diverges at the extremal points of temperature where its sign changes from $C_P>0$ to $C_P<0$, or vice versa, which leads to phase transitions, as shown in Figure \ref{C_P}. Compared with the ordinary Schwarzschild-AdS black hole with a finite $b$ parameter that has only one first-order phase transition, the self-regular Schwarzschild-AdS black hole has a complicated situation with one second-order, two first-order, and even no phase transitions,\footnote{The occurrence of different orders of phase transitions will be explained in the following context of this subsection.} depending on the values of vacuum pressure, see the bottom left diagram, the top right diagram, and the bottom right diagram, respectively, in Figure \ref{C_P}. Even for the case without the AdS background, i.e. $b\to \infty$, there is one first-order phase transition in the self-regular black hole, but no phase transition in the ordinary Schwarzschild black hole, see the top left diagram in Figure \ref{C_P}. In particular, one can see that the difference between the ordinary and self-regular black holes just exists in the near-extremal region, and such a difference disappears at a large horizon radius.

\begin{figure}[!ht]
\centering\includegraphics[height=7cm]{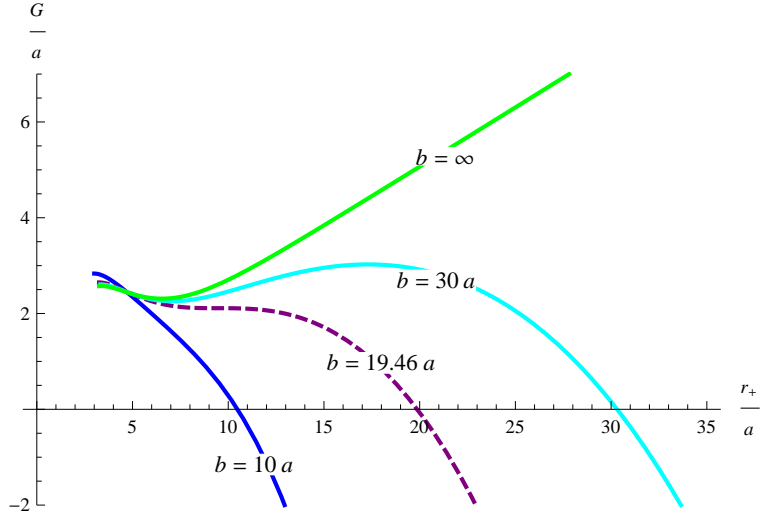}
\caption{Plot of the Gibbs free energy $G$ with respect to $r_+$ under different pressure. The green, cyan, and blue solid curves correspond to the different cases of ($b/a$, $G(r_0)/a$): ($\infty$, 2.575), (30, 2.607), and (10, 2.835), respectively. The dashed purple curve corresponds to the critical vacuum pressure with (19.46, 2.649).}
\label{fig:G_e}
\end{figure}

To analyze the phase transitions in detail, we should also investigate the thermodynamic potential. Since we are discussing an isothermal-isobaric ensemble, the thermodynamic potential should be the Gibbs free energy defined as
\begin{align}
G\equiv H-T_{\rm H} S=M-T_{\rm H} S,
\label{eq:G}
\end{align}
which is plotted in Figure \ref{fig:G_e}. By differentiating $G$ with respect to $r_+$, one obtains
\begin{eqnarray}
\frac{\partial G}{\partial r_+} &=& -S\left(\frac{\partial T_{\rm H}}{\partial r_+}\right),\\
\quad\quad \frac{\partial^2 G}{\partial r_+^2}&=& -\frac{C_P}{T_{\rm H}}\left(\frac{\partial T_{\rm H}}{\partial r_+}\right)^2-S \left(\frac{\partial^2 T_{\rm H}}{\partial r_+^2}\right),
\label{eq:G_d}
\end{eqnarray}
which implies that the curve of $G(r_+)$ has the same distributions of the extremal points and inflexion points as that of $T_{\rm H}(r_+)$ (see Figure \ref{fig:T_H_e}), apart from the extreme configuration with zero entropy. This property is quite helpful in our following analyses on phase transitions.

If we consider the case without the AdS background ($b \to \infty$), we can obtain from the green curve of Figure \ref{fig:T_H_e} that there are two phases of black holes with horizon radii $r_1$ and $r_2$,\footnote{$T_{max}$ is a maximum of temperature curve under some pressure, and $r_{max}$ is the corresponding radius of $T_{max}$; the similar notation applies to $T_{min}$ and $r_{min}$. Note that $r_{max} < r_{min}$, see Figure \ref{fig:T_H_e}.} the small  black hole $r_1<r_{max}$ and the large black hole $r_2>r_{max}$ in the region of temperature $0<T<T_{max}$, where $r_{max}=6.544a$ and $T_{max}=0.0096/a$, and no minimum of temperature in this case. For the large black hole with $r_2$, the negative heat capacity, see the top left diagram of Figure \ref{C_P}, implies that it is unstable. Further, our calculation shows that a large black hole has a higher Gibbs free energy than a small one. For instance, for a small black hole with $r_1=4.89a$, the Gibbs free energy is $2.42a$, but for a large black hole with $r_2=9.66a$, the Gibbs free energy equals $2.64a$ at the temperature $T=0.0080/a$. Thus, a large black hole would decay into a small one, i.e., the stable configuration with the positive capacity and lower energy.

In the case with the relatively low pressure ($19.46<b/a<\infty$), there are three regions divided by two vertical asymptotes in the $C_P-r_+$ plane (see, for instance, the top right diagram of Figure \ref{C_P}), corresponding to three types of black holes, or three phases, the small one with radius $r_1<r_{max}$ in the temperature range $T<T_{max}$, the medium one with $r_{max}<r_2<r_{min}$ in the range $T_{min}<T<T_{max}$, and the large one $r_3>r_{min}$ in the range $T>T_{min}$. Now we discuss different phases in different ranges of temperature. Due to the rich phase structures, this case is separated into the following three sub-cases.
\begin{itemize}
\item Taking $b/a=30$, for instance, we have $T_{max}=0.0113/a$ and $T_{min}=0.0092/a$, and their corresponding radii $r_{max}=7.070a$ and $T_{min}=17.31a$, respectively. For $T<T_{min}$, there is only a small black hole with $r_1<4.93a$ that is at least locally stable because of the positive heat capacity. Similarly, for $T>T_{max}$, only a large locally stable black hole with $r_3>33.67a$ exists. As to the temperature range $T_{min}<T<T_{max}$, the three types of black holes probably exist. Configurations of the small black hole with $r_1$ and of the large black hole with $r_3$ are at least locally stable and one of them with a lower Gibbs free energy is more stable; when the two configurations have the same Gibbs free energy, they can coexist and convert into each other, corresponding to a first-order phase transition since there is the entropy change during the process. The configuration of the medium black hole with the radius $r_2$ is unstable and probably decays into either the small black hole with $r_1$ or the large black hole with $r_3$, the more stable one with the lower Gibbs free energy.
\item When considering the pure thermal radiation with the vanishing Gibbs free energy~\cite{hawkingpage}, we can infer that the small black hole with $r_1$ in the temperature range $T<T_{max}$ is still locally stable since its Gibbs free energy is positive, and that the large black hole with $r_3$ can coexist with the pure thermal radiation at the Hawking-Page temperature $T_{\rm HP}=0.0107/a$. Moreover, if $T<T_{\rm HP}$, the small and large black holes with $r_1$ and $r_3$, respectively, in the range of temperature $T_{min}<T<T_{\rm HP}$, and the only small black hole with $r_1$ in the range of temperature $T<T_{min}$, have positive Gibbs free energy, so the pure thermal radiation is dominant. If $T>T_{\rm HP}$, in the range of $T_{\rm HP}<T<T_{max}$, the large black hole with $r_3$ has negative free energy and thus dominates, and in the range of $T>T_{max}$, the single large black hole is globally stable. Note that no black holes, even with positive Gibbs free energy, can decay into the pure thermal radiation because they cannot totally evaporate off, meaning that their radius decreases and finally reaches $r_0$ (the radius of extreme configurations) at which both the temperature and the heat capacity vanish. That is, the extreme black hole stops evaporating and its Gibbs free energy admits a local maximum,\footnote{Once a black hole forms, it would not disappear. To understand the final fate of black holes, we had better take a quantum view; that is, the extreme configuration corresponds to the ground state~\cite{1511.04865}, which cannot jump down into a lower state.} as argued in~\cite{1105.0188}.
It should be pointed out that the analyses for this and the above sub-case only apply to the range of pressure $b/a>27.53$ which leads to $T_{max}>T_{\rm HP}$.
\item  If the range of pressure is $19.46<b/a<27.53$ which leads to $T_{max}<T_{\rm HP}$, for the three types of black holes in the range of temperature $T_{min}<T<T_{max}$, the pure thermal radiation is dominant as all configurations of black holes have positive Gibbs free energy. Moreover, in the range of temperature $T_{max}<T<T_{\rm HP}$, the pure thermal radiation is still dominant because the single large black hole with horizon radius $r_3$ has positive Gibbs free energy. At last, if $T>T_{\rm HP}$, the single large black hole is globally stable.
\end{itemize}

For the critical pressure case ($b/a=19.46$), the two vertical asymptotes in $C_P-r_+$ plane merge into one and the second region with negative heat capacity disappears (see the bottom left diagram of Figure \ref{C_P}). Only one locally stable small black hole ($r_+<r_c=9.47a$) or one at least locally stable large black hole ($r_+>r_c=9.47a$) exists. The phase transition at the inflexion is second order because the entropy is continuous in the two phases. Here are also three situations: ($\mathrm{i}$) If $r_+<r_{\rm HP}$, where $r_{\rm HP}=19.87a$ is the horizon radius when the black hole is at the Hawking-Page temperature, the black hole with positive free energy stays in a locally stable configuration; ($\mathrm{ii}$) If $r_+=r_{\rm HP}$, the black hole with vanishing free energy coexists with the pure thermal radiation; ($\mathrm{iii}$) If $r_+>r_{\rm HP}$, the black hole with negative free energy is globally stable.

If the ensemble is in the relatively high pressure ($b/a<19.46$), the heat capacity changes continuously with respect to the radius without divergence (see the bottom right diagram of Figure \ref{C_P}). As a result, there is no phase transition between black holes. The analysis about the Hawking-Page phase transition resembles that of the case at the critical pressure.

We would like to point out that a recent work by Frassino, K\"oppel, and Nicolini reveals the phase structure of the holographic metric~\cite{1604.03263}. This phase structure is similar to that of our model, where the reason will be given in section 4. In addition,  the minimal length is treated as a variable in ref.~\cite{1105.0188}, where the increase of the minimal length leads to the same tendency of phase transitions as that of the decrease of the AdS radius we have just discussed above. We note that such a comparability coincides with our analysis in section 2 where the minimal length is related to the equivalent dS negative pressure, see eq.~(\ref{dSmetric}).

\subsection{Thermodynamics of the collapsed shell model}
\begin{figure}[!ht]
\centering\includegraphics[height=7cm]{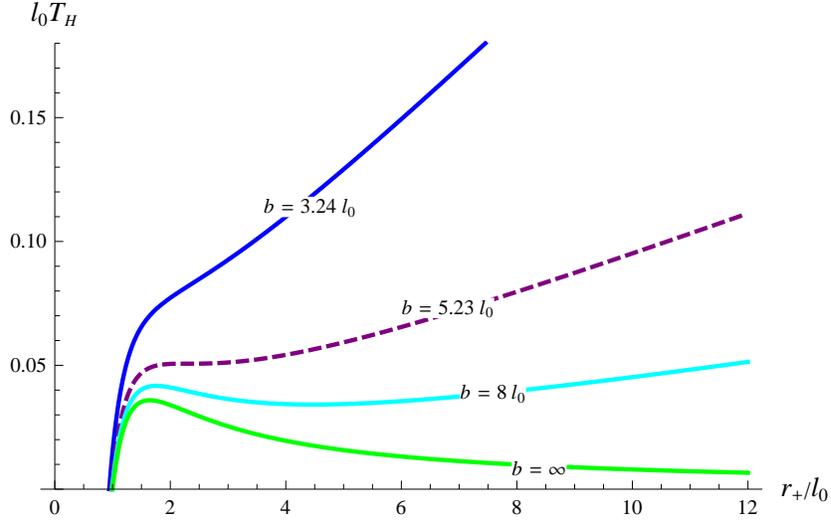}
\caption{Plot of the Hawking temperature $T_{\rm H}$ with respect to $r_+$ under different vacuum pressure. Particularly, the dashed purple curve corresponds to the case of the critical pressure.}
\label{T_H_s}
\end{figure}

\begin{figure}[!ht]
\centering\includegraphics[height=7cm]{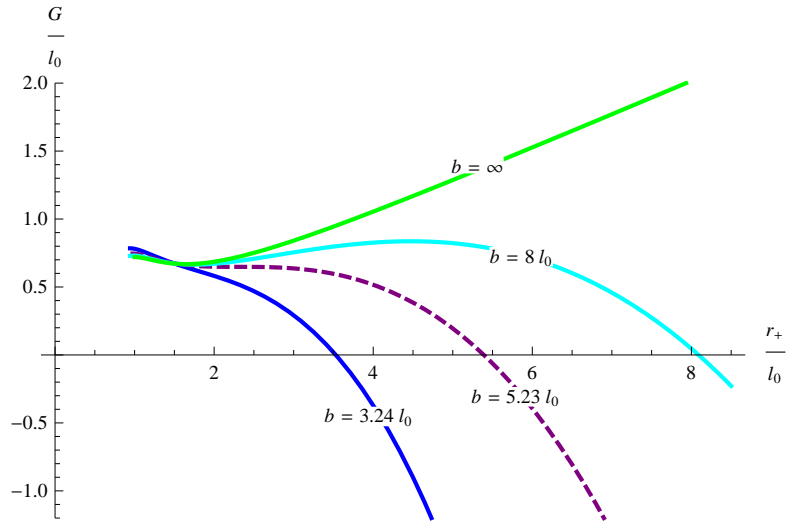}
\caption{Plot of the Gibbs free energy $G$ with respect to $r_+$ under different pressure. Particularly, the dashed purple curve corresponds to the case of the critical pressure.}
\label{G_s}
\end{figure}

We at first derive the Hawking temperature of the collapsed shell  from eq.~(\ref{eq:T_H(kappa)}),
\begin{align}
T_{\rm H}=\frac{5 b^2 \left(r_+^3-l_0^3\right)+3 r_+^2 \left(5 r_+^3-l_0^3\right)}{4 \pi  b^2 r_+ \left(l_0^3+5 r_+^3\right)},
\end{align}
which is plotted in Figure \ref{T_H_s}.
Then we calculate the entropy,
\begin{align}
S= \pi \left(r_+^2-r_0^2 \right)-\frac{4}{5}\pi l_0^3 \left(\frac{1}{r_+}-\frac{1}{r_0}\right)-\frac{1}{50} \pi l_0^6 \left(\frac{1}{r_+^4}-\frac{1}{r_0^4}\right),
\label{eq:S_s}
\end{align}
where $r_0$, the horizon radius of the extreme black hole, is the solution of $\frac{\mathrm{d}M}{\mathrm{d}r_{\rm H}}=0$, cf. eq.~(\ref{eq:M}). The first term which equals one fourth of the horizon area satisfies the area law of the ordinary black hole thermodynamics, and the other terms are quantum corrections which are induced by the minimal length. We notice that Figure \ref{T_H_s} presents similar features to that in Figure \ref{fig:T_H_e}, which implies that the thermodynamic properties of the collapsed shell model resemble that of the hydrogen-atom-like model due to the relations between the temperature and the heat capacity and between the temperature and the Gibbs free energy, see  eqs.~(\ref{eq:C_P}) and (\ref{eq:G_d}).
 From the physical point of view, the equilibrium state is described by the equation of state plotted in
Figure \ref{T_H_s} with isobars in the temperature-horizon plane, which displays the thermodynamic similarities for the two models, cf. Figure \ref{T_H_s} and Figure \ref{fig:T_H_e}. In addition, in order to consider the pure thermal radiation of the AdS background, we have to investigate the Gibbs free energy. To this end, the $G-r_+$ plane is plotted in Figure \ref{G_s}, which shows that the collapsed shell model indeed has the similar phase structure to that of the hydrogen-atom-like model, see Figure \ref{fig:G_e}. Thus, the remaining  discussions are not necessary to be repeated.

\section{General analysis}
Our discussions in the above two sections are based on two specific models which show the similar thermodynamic properties to that of the other models constructed in refs.~\cite{1212.5044, 1110.5332, 1410.1706}, where those mass distributions can also be understood as a kind of step functions with continuity. In this section, we demonstrate that the models with such a mass distribution have similar thermodynamic properties in a general way.

\begin{figure}[!ht]
\centering\includegraphics[height=7cm]{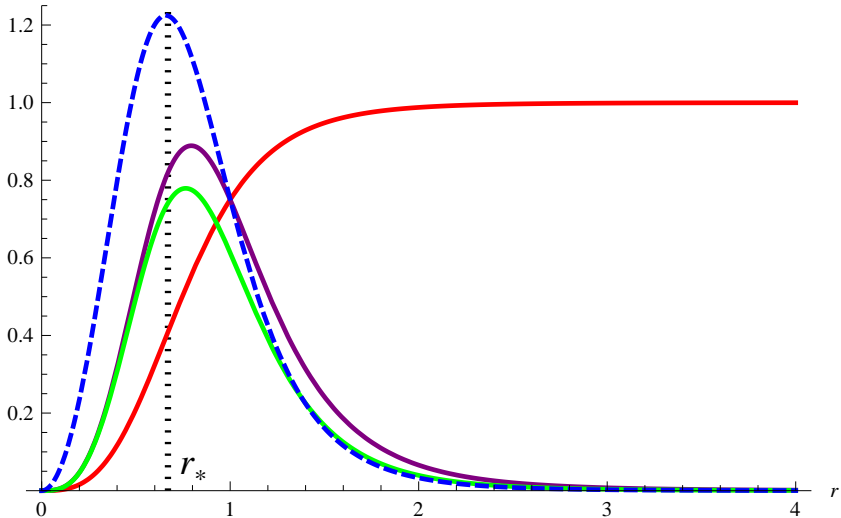}
\caption{ Sketches of $f(r)$ (red), $f'(r)r$ (purple), and $f'(r)r\left(1-\frac{2r^2}{b^2+3r^2}\right)$ (green). The curve of $f'(r)$, the dashed blue curve with $f''(r_*)=0$, is drawn for reference.}
\label{G_r}
\end{figure}

When the mass density of black holes takes some analytic form of the $\delta$-function with the minimal length as a parameter, the metric eq.~(\ref{eq:ds^2_ads}) can be extended to a general self-regular Schwarzschild-AdS solution of the modified Einstein's equations \cite{0510112},
\begin{align}
ds^2=-\left(1-\frac{2 f(r)M}{r}+\frac{r^2}{b^2}\right)\mathrm{d}t^2+\left(1-\frac{2 f(r)M}{r}+
\frac{r^2}{b^2}\right)^{-1}\mathrm{d}r^2+r^2\,\mathrm{d}\Omega^2,
\label{ds^2}
\end{align}
where $f(r)$, the original $\mathcal{M}(r)/M$ in eq.~(\ref{eq:ds^2_ads}), is unfixed but is supposed to be a kind of step functions with continuity, see the red curve in Figure \ref{G_r} for illustration. Specifically, as $\rho(r)=\frac{M}{4\pi r^2}\frac{\mathrm{d}f}{\mathrm{d}r}$ is finite everywhere, $f(r)$ is a third-order infinitesimal with respect to $r$ for the ``collapsed core" (see, for instance, our first model) or a higher order one for the ``collapsed shell" (see, for instance, our second model) near the origin in order to get rid of the singularity of the metric, and it approaches one at a large $r$.\footnote{Here $f(r)$ can also be dealt with as an efficient gravitational coupling, or the so-called ``running" gravitational coupling which describes an asymptotically free gravitational model~\cite{1212.5044}.} The relationship between the total mass $M$ and the horizon radius $r_{\rm H}$ can be obtained as usual,
\begin{align}
M=\frac{r_{\rm H}}{2 f(r_{\rm H})}\left(1+\frac{r_{\rm H}^2}{b^2}\right).
\label{eq:M_G}
\end{align}
Since $M$ approaches the positive infinity both at $r_{\rm H}\rightarrow 0$ and $r_{\rm H}\rightarrow\infty$, there is at least one minimum point. Based on the equation,
\begin{align}
\left.\frac{\mathrm{d}M}{\mathrm{d}r_{\rm H}}\right|_{r_{\rm H}=r_0}=\left.\frac{f(r_{\rm H})(b^2+3r_{\rm H}^2)-f'(r_{\rm H})r_{\rm H}(b^2+r_{\rm H}^2)}{2 f(r_{\rm H})^2 b^2}\right|_{r_{\rm H}=r_0}=0,
\label{eq:M_G}
\end{align}
we obtain  the extremal point $r_0$ satisfying $f(r_0)=f^{\prime}(r_0)r_0\left(1-\frac{2r_0^2}{b^2+3r_0^2}\right)$. Figure \ref{G_r} illustrates that there is only one intersection for $f(r_{\rm H})$ and $f'(r_{\rm H})r_{\rm H}\left(1-\frac{2r_{\rm H}^2}{b^2+3r_{\rm H}^2}\right)$, and thus there exists only one extreme black hole with the minimal mass and the minimal radius.

The existence of extreme black holes for the metric solution eq.~(\ref{ds^2}) can be shown by the following analysis. For the case without the AdS background, i.e., $b \to \infty$, we obtain\footnote{The subscript of horizon radii is omitted for the sake of convenience hereafter.} $\frac{\mathrm{d}M}{\mathrm{d}r}=\frac{f(r)-f'(r)r }{2 f(r)^2}$. The sketches of $f(r)$ and $f'(r)r$ are drawn in Figure \ref{G_r}. For a small $r$, as $f(r)$ is at least a third-order infinitesimal with respect to $r$, we obtain $f'(r)r>\frac{f(r)-f(0)}{r}r=f(r)$, that is, the curve of $f'(r)r$ is over that of $f(r)$. Next, in order to analyze the increasing tendency of the two functions, we calculate their derivatives, $[f'(r)r]'=f''(r)r+f'(r)$ and $f'(r)$, and deduce $[f'(r)r]'>f'(r)$ if $r<r_*$, see the caption of Figure  \ref{G_r} for the definition of $r_*$. As a result, only if $r>r_*$ can one obtain $[f'(r)r]'<f'(r)$, which leads to  intersection of the two curves. After the two curves intersect at the horizon radius\footnote{Here we have normalized the coordinate of the intersection, which will be inferred to be the minimal length.}  $r=1$, $f''(r)<0$ renders $[f'(r)r]'<f'(r)$, which implies that $r=1$ is the single point where $f(r)-f'(r)r$ vanishes, i.e., the only one extreme configuration of black holes. Moreover, for the case with the AdS background, i.e., the parameter $b$ is finite and satisfies the hoop conjecture,  the intersection of $f(r)$ and $f'(r)r\left(1-\frac{2r^2}{b^2+3r^2}\right)$ will have a shift to the left if compared with the case without the AdS background, i.e., the intersecting point is smaller than $1$ because $f'(r)r\left(1-\frac{2r^2}{b^2+3r^2}\right)$ is smaller than $f'(r)r$.

 %Apart from phase transitions between black holes and the pure thermal radiation,
The thermodynamic properties of black holes merely depend on the equation of state, so we just need to analyze the relationship between the Hawking temperature $T_{\rm H}$ and the specific volume, or simply between the Hawking temperature and the outer horizon radius $r_+$, see Footnote $\ref{2}$. From eq.~$(\ref{eq:T_H(kappa)})$, we calculate the temperature,
\begin{eqnarray}
T_{\rm H}=\frac{1}{4\pi}\frac{f(r_+)(b^2+3r_+^2)-f'(r_+)(b^2r_++r_+^3)}{f(r_+)b^2 r_+}=\frac{f(r_+)}{2\pi r_+}\frac{\mathrm{d}M}{\mathrm{d}r_+},
\label{eq:T_G}
\end{eqnarray}
which vanishes at the extremal radius $r_0$ that is the solution of $\frac{\mathrm{d}M}{\mathrm{d}r_+}=0$, and is always positive at other radii.

\subsection{Features of temperature of black holes without the AdS background}
In the absence of the AdS background ($b \to \infty$), the Hawking temperature and its derivative with respect to $r_+$ take the forms,
\begin{eqnarray}
T_{\rm H}&=&\frac{1}{4\pi}\left(\frac{1}{r_+}-\frac{f'(r_+)}{f(r_+)}\right),  \\
\frac{\mathrm{d}T_{\rm H}}{\mathrm{d}r_+}&=&-\frac{1}{4\pi}\frac{f(r_+)^2-f'(r_+)^2 r_+^2+f''(r_+)f(r_+)r_+^2}{f(r_+)^2r_+^2},\label{maxT}
\end{eqnarray}
and the temperature vanishes at $r_{+}\rightarrow\infty$. As to the derivative of the temperature, it is positive due to  $f(r_0)=f^{\prime}(r_0)r_0$ and $f''(r_0)<0$ at the extremal radius $r_0$ and close to zero at infinity. As a result, there is at least one local maximum temperature $T_1$ whose corresponding radius $r_1$ satisfies the condition from the vanishing of eq.~(\ref{maxT}) at $r_+=r_1$,
\begin{align}
f''(r_1)f(r_1)r_1^2 -f'(r_1)^2 r_1^2+f(r_1)^2=0.
\label{eq:T'_0}
\end{align}
In the following we prove that $T_1$ is the global maximum and thus indicate the similar $T_{\rm H}-r_+$ relation to that of the above two models depicted by Figures \ref{fig:T_H_e} and \ref{T_H_s}, which ensures that the thermodynamic properties of the general case are similar to that of the other specific models mentioned.

\begin{figure}[!ht]
\centering\includegraphics[height=7cm]{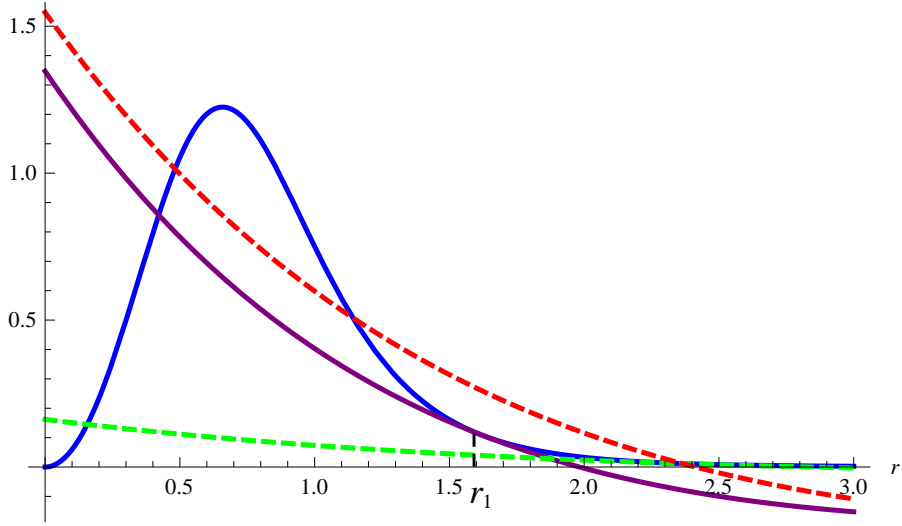}
\caption{ Sketches of $f'(r)$ (blue) and $g'(r)=c_1(1+c_2 r) \exp({c_2 r})$ (purple). The dashed red curve and the dashed green curve represent ${\tilde g}'(r)={\tilde c}_1(1+{\tilde c}_2 r) \exp({{\tilde c}_2 r})$ with ${\tilde c}_1>c_1$ and ${\tilde c}_1<c_1$, respectively.}
\label{fig:g_r}
\end{figure}

Set function $g(r)$ satisfy the following second-order differential equation,
\begin{align}
g''(r)g(r)r^2 -g'(r)^2 r^2+g(r)^2=0,
\label{eq:g_r}
\end{align}
we have a general solution, $g(r)=c_1 r \exp({c_2 r})$, where $c_1$ and $c_2$ are two parameters to be determined. By choosing a positive parameter $c_1=\frac{f(r_1)}{r_1}\exp[-f'(r_1)r_1+f(r_1)]$ and a negative parameter $c_2=\frac{f'(r_1)r_1-f(r_1)}{r_1}$, we obtain the particular solution that ensures $g(r_1)=f(r_1)$ and $g'(r_1)=f'(r_1)$. Again using $g(r_1)=f(r_1)$, $g'(r_1)=f'(r_1)$, and eqs.~(\ref{eq:T'_0}) and (\ref{eq:g_r}), we deduce $g''(r_1)=f''(r_1)$. Since it is difficult to depict the equality of the second derivatives, $g''(r_1)=f''(r_1)$, on the plot of curves of $f(r)$ and $g(r)$, we choose to draw the curves of $f'(r)$ and $g'(r)$ instead, see Figure \ref{fig:g_r}. One can infer that the two curves are tangent at $r_1$ due to $g'(r_1)=f'(r_1)$ and $g''(r_1)=f''(r_1)$. In addition, we note that as a result of $g(r)=\int_{0}^{r}g'(r')\mathrm{d}r'+g(0)$ and $g(0)=0$, $g(r)$ is equal to the integral area surrounded by the curve $g'(r)$ and the transverse axis from the origin to $r$, and the same analysis is applicable for $f(r)$. Consequently, $g(r_1)=f(r_1)$ indicates that the  integral areas surrounded by the curve $g'(r)$ and the transverse interval $[0, r_1]$ and by the curve $f'(r)$ and the transverse interval $[0, r_1]$ are same,  see the two solid curves in Figure \ref{fig:g_r}. That is to say, the equality of the two integral areas implies the existence of the maximum of temperature in the $T_{\rm H}-r_+$ plane.

Assuming that there still exists one local minimum of temperature ${\tilde T}_1$ at ${\tilde r}_1$ on the curve of $T_{\rm H}$ that meets $f''({\tilde r}_1)f({\tilde r}_1){\tilde r}_1^2-f'({\tilde r}_1)^2 {\tilde r}_1^2+f({\tilde r}_1)^2=0$, we can certainly find out another particular solution ${\tilde g}(r)={\tilde c}_1 r \exp({{\tilde c}_2 r})$ with ${\tilde c}_1=\frac{f({\tilde r}_1)}{{\tilde r}_1}\exp[-f'({\tilde r}_1){\tilde r}_1+f({\tilde r}_1)]$  and ${\tilde c}_2=\frac{f'({\tilde r}_1){\tilde r}_1-f({\tilde r}_1)}{{\tilde r}_1}$ to the differential equation ${\tilde g}''(r){\tilde g}(r)r^2-{\tilde g}'(r)^2 r^2+{\tilde g}(r)^2=0$. This particular solution satisfies ${\tilde g}({\tilde r}_1)=f({\tilde r}_1)$, ${\tilde g}'({\tilde r}_1)=f'({\tilde r}_1)$, and ${\tilde g}''({\tilde r}_1)=f''({\tilde r}_1)$. Using eq.~(\ref{maxT}), we have the maximum of temperature at $r_1$,
\begin{align}
T_{\rm H}(r_1)=\frac{1}{4\pi}\left(\frac{1}{r_1}-\frac{f'(r_1)}{f(r_1)}\right)=\frac{1}{4\pi}\left(\frac{1}{r_1}
-\frac{g'(r_1)}{g(r_1)}\right)=-\frac{1}{4\pi}c_2>0,
\label{eq:T_r_e}
\end{align}
which means that $T_{\rm H}(r_1)$ is only related to the negative parameter $c_2$. Considering $T_{\rm H}(r_1)>T_{\rm H}({\tilde r}_1)$, one can infer ${\tilde c}_2>c_2$. Next we compare $g'(r)$ with ${\tilde g}'(r)={\tilde c}_1(1+{\tilde c}_2 r) \exp({{\tilde c}_2 r})$ by considering the following two cases in order to determine that the local minimum of temperature ${\tilde T}_1$ conjectured above does not exist.

(i) For the case ${\tilde c}_1>c_1$, we can get ${\tilde g}'(r)>g'(r)$, which means that the curve of ${\tilde g}'(r)$ is always above that of $g'(r)$. So the integral area surrounded by ${\tilde g}'(r)$ and the interval $[0, \tilde r_1]$ is not same as that surrounded by $f'(r)$ and the interval $[0, \tilde r_1]$. In fact, we always have $\tilde g(\tilde r_1)>f(\tilde r_1)$ in this case, see, for instance, the sketches in Figure \ref{fig:g_r}.

(ii) For the case ${\tilde c}_1<c_1$, we can first get ${\tilde g}'(0)={\tilde c}_1<c_1=g'(0)$; then if the two curves of ${\tilde g}'(r)$ and $f'(r)$ are tangent at ${\tilde r}_1$, we can infer ${\tilde g}({\tilde r}_1)=\int^{{\tilde r}_1}_0 {\tilde g}'(r) \mathrm{d}r<\int^{{\tilde r}_1}_0 f'(r) \mathrm{d}r =f({\tilde r}_1)$, which also leads to the inequality of the integral areas, see the sketches in Figure \ref{fig:g_r}.

In a word, no suitable parameters $\tilde c_1$ and $\tilde c_2$ can be found to simultaneously meet ${\tilde g}({\tilde r}_1)=f({\tilde r}_1)$, ${\tilde g}'({\tilde r}_1)=f'({\tilde r}_1)$, and ${\tilde g}''({\tilde r}_1)=f''({\tilde r}_1)$, or in other words, to meet $T_{\rm H}'({\tilde r}_1)=0$. As a result, the curve of $T_{\rm H}$ without the AdS background has one and only one extremal point, i.e, the maximum point, which certainly leads to the conclusion that the $T_{\rm H}-r_+$ relation in this  case is similar to the green curves of Figures \ref{fig:T_H_e} and \ref{T_H_s}.

\subsection{Features of temperature of black holes with the AdS background}
For the case with the AdS background, we rewrite the Hawking temperature eq.~(\ref{eq:T_G}) and calculate the corresponding derivative as follows:
\begin{eqnarray}
%\begin{split}
T_{\rm H}&=&\frac{1}{4\pi}\left(\frac{f(r_+)-f'(r_+)r_+}{f(r_+) r_+}+\frac{3 f(r_+)r_+-f'(r_+)r_+^2}{f(r_+)b^2}\right),
\label{T_H_b}\\
\frac{\mathrm{d}T_{\rm H}}{\mathrm{d}r_+}
%&=&\frac{1}{4\pi}\frac{r_+^2 \left(b^2+r_+^2\right)f'(r_+)^2-r_+^2 f(r_+) \left(\left(b^2+r_+^2\right) f''(r_+)+2 r_+ f'(r_+)\right)+\left(3 r_+^2-b^2\right) f(r_+)^2}{b^2 r_+^2 f(r_+)^2}  \nonumber\\
&=&\frac{1}{4\pi}\left[-\frac{f(r_+)^2-f'(r_+)^2 r_+^2+f''(r_+)f(r_+)r_+^2}{f(r_+)^2r_+^2}\right.  \nonumber\\
& &\;\;\;\;\;\;\;\left.+\frac{3 f(r_+)^2+f'(r_+)^2 r_+^2-2f(r_+)f'(r_+) r_+-f(r_+)f''(r_+)r_+^2}{b^2 f(r_+)^2}\right].\label{T'_G}
%\end{split}
\end{eqnarray}

The Hawking temperature is still zero for extreme black holes, but because of the finite $b$ parameter, it grows to infinity as the horizon radius increases. As to the derivative of the temperature, $T'_{\rm H}$, it is positive according to  eq.~(\ref{eq:T_G}) at the extremal radius $r_0$ and also positive at infinity because it goes to $3/(4\pi b^2)$. For a relatively large $b$, the zero point of $T_{\rm H}'$, where the main part of $T_{\rm H}'$ is the first term of eq.~(\ref{T'_G}), will slightly shift to the right compared with the case of $b \to \infty$, and $T_{\rm H}'$ can also be negative at some radius. As a result, there exist at least one maximum and one following minimum on the curve of $T_{\rm H}$ when the horizon radius increases. In this case, the two points can be proven to be the only pair of extremal points of $T_{\rm H}$ where the numerator of $T'_{\rm H}$ vanishes. In fact, to the differential equation,\footnote{This equation comes from the condition $\frac{\mathrm{d}T_{\rm H}}{\mathrm{d}r_+}=0$.}
\begin{align}
h(r)h''(r)r^2(b^2+r^2)-h'(r)^2 r^2(b^2+r^2)+2 h(r)h'(r)r^3 -h(r)^2(3r^2-b^2)=0,
\end{align}
we can find a general solution,
\begin{align}
h(r)=\lambda r(r^2+b^2) \exp\left({\frac{\beta}{b}\tan^{-1}{\frac{r}{b}}}\right),
\end{align}
where $\lambda$ and $\beta$ are two parameters to be determined, and give its derivative as follows,
\begin{align}
h'(r)=\lambda (3r^2+{\beta} r+b^2) \exp\left({\frac{\beta}{b}\tan^{-1}{\frac{r}{b}}}\right).
\end{align}

We can obtain two particular solutions, $h_1(r)=\lambda_1 r(r^2+b^2) \exp\left({\frac{\beta_1}{b}\tan^{-1}{\frac{r}{b}}}\right)$ and $h_2(r)=\lambda_2 r(r^2+b^2) \exp\left({\frac{\beta_2}{b}\tan^{-1}{\frac{r}{b}}}\right)$,  where the two pairs of parameters take the forms,\footnote{The positivity of the $\lambda$ parameter and the negativity of the $\beta$ parameter can be determined by eq.~(\ref{eq:M_G}).}
\begin{eqnarray}
\lambda_1&=&\frac{f(r_{_{\rm{\uppercase\expandafter{\romannumeral1}}}})}{r_{_{\rm{\uppercase\expandafter{\romannumeral1}}}}
(r_{_{\rm{\uppercase\expandafter{\romannumeral1}}}}^2+b^2)}\exp\left(-\frac{\beta_1}{b}\tan^{-1}\frac{r_{_{\rm{\uppercase\expandafter{\romannumeral1}}}}}{b}\right) >0,\nonumber \\
\beta_1&=&\frac{f'(r_{_{\rm{\uppercase\expandafter{\romannumeral1}}}})r_{_{\rm{\uppercase\expandafter{\romannumeral1}}}}(r_{_{\rm{\uppercase\expandafter{\romannumeral1}}}}^2+b^2)-f(r_{_{\rm{\uppercase\expandafter{\romannumeral1}}}})(3r_{_{\rm{\uppercase\expandafter{\romannumeral1}}}}^2
+b^2)}{f(r_{_{\rm{\uppercase\expandafter{\romannumeral1}}}})r_{_{\rm{\uppercase\expandafter{\romannumeral1}}}}}<0;\nonumber \\
\lambda_2&=&\frac{f(r_{_{\rm{\uppercase\expandafter{\romannumeral2}}}})}{r_{_{\rm{\uppercase\expandafter{\romannumeral2}}}}(r_{_{\rm{\uppercase\expandafter{\romannumeral2}}}}^2+b^2)
}\exp\left(-\frac{\beta_2}{b}\tan^{-1}\frac{r_{_{\rm{\uppercase\expandafter{\romannumeral2}}}}}{b}\right)>0, \nonumber \\
\beta_2&=&\frac{f'(r_{_{\rm{\uppercase\expandafter{\romannumeral2}}}})r_{_{\rm{\uppercase\expandafter{\romannumeral2}}}}(r_{_{\rm{\uppercase\expandafter{\romannumeral2}}}}^2+b^2)-f(r_{_{\rm{\uppercase\expandafter{\romannumeral2}}}})(3r_{_{\rm{\uppercase\expandafter{\romannumeral2}}}}^2+b^2)}
{f(r_{_{\rm{\uppercase\expandafter{\romannumeral2}}}})
r_{_{\rm{\uppercase\expandafter{\romannumeral2}}}}}<0, \nonumber
\end{eqnarray}
and $r_{_{\rm{\uppercase\expandafter{\romannumeral1}}}}$ and $r_{_{\rm{\uppercase\expandafter{\romannumeral2}}}}$ are the maximum and minimum points, respectively, and meet the inequality $r_{_{\rm{\uppercase\expandafter{\romannumeral1}}}}<r_{_{\rm{\uppercase\expandafter{\romannumeral2}}}}$.
The particular solutions satisfy the conditions,
$h_1(r_{_{\rm{\uppercase\expandafter{\romannumeral1}}}})=f(r_{_{\rm{\uppercase\expandafter{\romannumeral1}}}})$,
$h_1'(r_{_{\rm{\uppercase\expandafter{\romannumeral1}}}})=f'(r_{_{\rm{\uppercase\expandafter{\romannumeral1}}}})$,  and
$h_1''(r_{_{\rm{\uppercase\expandafter{\romannumeral1}}}})=f''(r_{_{\rm{\uppercase\expandafter{\romannumeral1}}}})$;
$h_2(r_{_{\rm{\uppercase\expandafter{\romannumeral2}}}})=f(r_{_{\rm{\uppercase\expandafter{\romannumeral2}}}})$,
$h_2'(r_{_{\rm{\uppercase\expandafter{\romannumeral2}}}})=f'(r_{_{\rm{\uppercase\expandafter{\romannumeral2}}}})$, and
$h_2''(r_{_{\rm{\uppercase\expandafter{\romannumeral2}}}})=f''(r_{_{\rm{\uppercase\expandafter{\romannumeral2}}}})$.
As a result, the curve of $f'(r)$ is tangent to that of $h_1'(r)$ at $r_{_{\rm{\uppercase\expandafter{\romannumeral1}}}}$ and also tangent to that of $h_2'(r)$ at  $r_{_{\rm{\uppercase\expandafter{\romannumeral2}}}}$. Again considering $h(r)=\int_0^r h'(\tilde r)\mathrm{d}{\tilde r}$ and $h(0)=0$, we deduce that the integral area surrounded by the curve $h_1'(r)$ and the transverse interval $[0, r_{_{\rm{\uppercase\expandafter{\romannumeral1}}}}]$ equals that by the curve $f'(r)$ and the transverse interval $[0, r_{_{\rm{\uppercase\expandafter{\romannumeral1}}}}]$, and the same result for the integral areas surrounded by the curve $h_2'(r)$ and transverse interval $[0, r_{_{\rm{\uppercase\expandafter{\romannumeral2}}}}]$ and by the curve $f'(r)$ and the transverse interval $[0, r_{_{\rm{\uppercase\expandafter{\romannumeral2}}}}]$.

Using eq.~(\ref{T_H_b}), we have the temperature at the extremal (maximum and minimum) points $r_e$,
\begin{align}
T_{\rm H}(r_e)=\frac{1}{4\pi}\left[\frac{h(r_e)-h'(r_e)r_e}{h(r_e)r_e}+\frac{3 h(r_e)r_e-h'(r_e)r_e^2}{h(r_e)b^2}\right]=-\frac{\beta}{4 \pi b^2},
\end{align}
which is only related to the negative parameter $\beta$. Due to $T_{\rm H}(r_{_{\rm{\uppercase\expandafter{\romannumeral1}}}})>T_{\rm H}(r_{_{\rm{\uppercase\expandafter{\romannumeral2}}}})$, we obtain $\beta_1<\beta_2$. As to the relationship between $\lambda_1$ and $\lambda_2$, if $\lambda_1<\lambda_2$ is taken, the curve of $h_1(r)$ is always below that of $h_2(r)$, which cannot lead to $h_1(r_{_{\rm{\uppercase\expandafter{\romannumeral1}}}})=f(r_{_{\rm{\uppercase\expandafter{\romannumeral1}}}})$ and $h_2(r_{_{\rm{\uppercase\expandafter{\romannumeral2}}}})=f(r_{_{\rm{\uppercase\expandafter{\romannumeral2}}}})$ simultaneously. As a result, we get $\lambda_1>\lambda_2$.
%Taking $h'(0)=\lambda b^2$ and $r_{_{\rm{\uppercase\expandafter{\romannumeral1}}}}<r_{_{\rm{\uppercase\expandafter{\romannumeral2}}}}$ into consideration, we

\begin{figure}[!ht]
\centering\includegraphics[width=14cm,height=14cm]{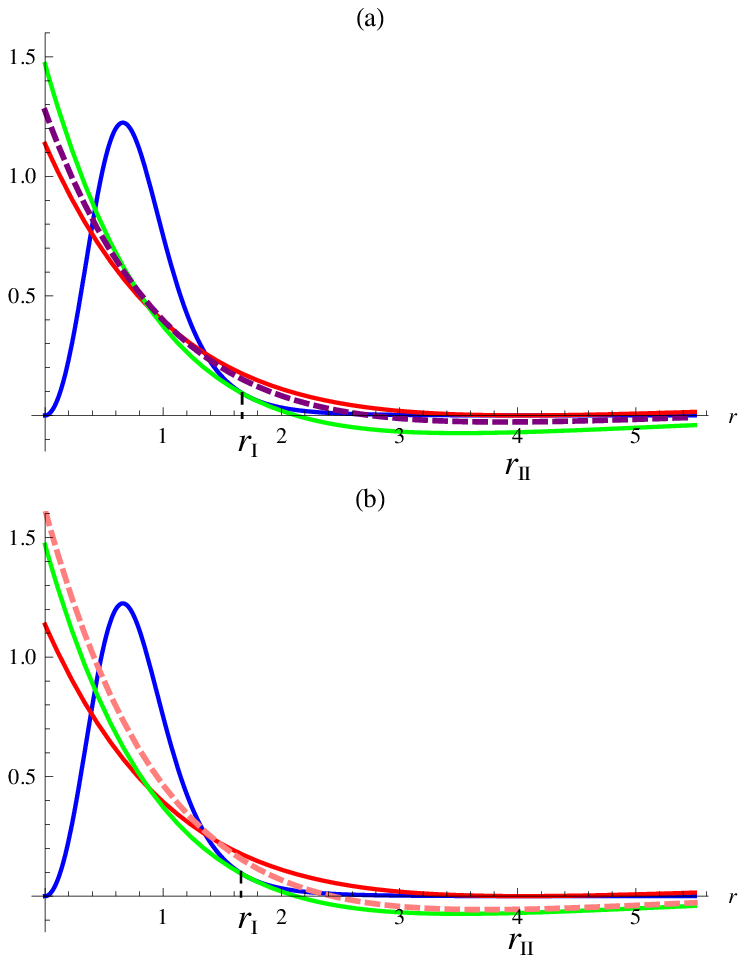}
\caption{Sketches of $f'(r)$ (blue), $h_1'(r)=\lambda_1 (3r^2+\beta_1 r+b^2)\exp(\frac{\beta_1}{b}\tan^{-1}{\frac{r}{b}})$ (green), and $h_2'(r)=\lambda_2 (3r^2+\beta_2 r+b^2)\exp(\frac{\beta_2}{b}\tan^{-1}{\frac{r}{b}})$ (red). The dashed purple (a) and dashed pink (b) curves represent $\tilde h'(r)=\tilde\lambda (3r^2+\tilde\beta r+b^2)\exp(\frac{\tilde\beta}{b}\tan^{-1}{\frac{r}{b}})$ with the different ranges of parameters, ($\beta_1<\tilde\beta<\beta_2$, $\lambda_2<\tilde\lambda<\lambda_1$) and ($\beta_1<\tilde\beta<\beta_2$, $\lambda_2<\lambda_1<\tilde\lambda$), respectively.}
\label{h_r_1}
\end{figure}

\begin{figure}[!ht]
\centering\includegraphics[width=14cm,height=14cm]{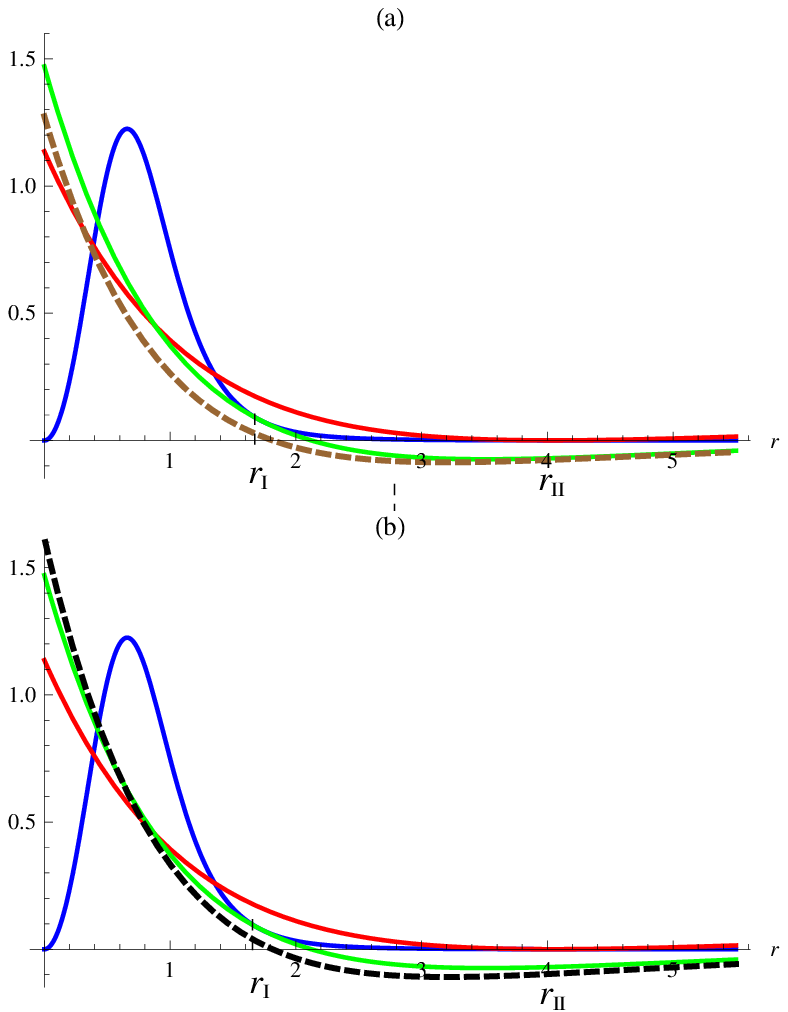}
\caption{Sketches of $f'(r)$ (blue), $h_1'(r)=\lambda_1 (3r^2+\beta_1 r+b^2)\exp(\frac{\beta_1}{b}\tan^{-1}{\frac{r}{b}})$ (green), and $h_2'(r)=\lambda_2 (3r^2+\beta_2 r+b^2)\exp(\frac{\beta_2}{b}\tan^{-1}{\frac{r}{b}})$ (red). The dashed brown (a) and dashed black (b) curves represent $\tilde h'(r)=\tilde\lambda (3r^2+\tilde\beta r+b^2)\exp(\frac{\tilde\beta}{b}\tan^{-1}{\frac{r}{b}})$ with the different ranges of parameters, ($\tilde\beta<\beta_1<\beta_2$, $\lambda_2<\tilde\lambda<\lambda_1$) and ($\tilde\beta<\beta_1<\beta_2$, $\lambda_2<\lambda_1<\tilde\lambda$), respectively.}
\label{h_r_2}
\end{figure}

If another maximum point $\tilde r_{_{\rm{\uppercase\expandafter{\romannumeral1}}}}$ following $r_{_{\rm{\uppercase\expandafter{\romannumeral2}}}}$ ($\tilde r_{_{\rm{\uppercase\expandafter{\romannumeral1}}}}>r_{_{\rm{\uppercase\expandafter{\romannumeral2}}}}$) is conjectured on the curve of $T_{\rm H}$, one can seek out the corresponding particular solution, $\tilde h(r)=\tilde\lambda r(r^2+b^2) \exp\left({\frac{\tilde\beta}{b}\tan^{-1}{\frac{r}{b}}}\right)$, where the parameters take the forms,
\begin{eqnarray}
\tilde\lambda &=& \frac{f(\tilde r_{_{\rm{\uppercase\expandafter{\romannumeral1}}}})}{\tilde r_{_{\rm{\uppercase\expandafter{\romannumeral1}}}}
(\tilde r_{_{\rm{\uppercase\expandafter{\romannumeral1}}}}^2+b^2)}\exp\left(-\frac{\tilde \beta_1}{b}\tan^{-1}\frac{\tilde r_{_{\rm{\uppercase\expandafter{\romannumeral1}}}}}{b}\right) >0, \nonumber \\
\tilde\beta &=& \frac{f'(\tilde r_{_{\rm{\uppercase\expandafter{\romannumeral1}}}})(\tilde r_{_{\rm{\uppercase\expandafter{\romannumeral1}}}}^3+b^2
\tilde r_{_{\rm{\uppercase\expandafter{\romannumeral1}}}})-f(\tilde r_{_{\rm{\uppercase\expandafter{\romannumeral1}}}})(3\tilde r_{_{\rm{\uppercase\expandafter{\romannumeral1}}}}^2
+b^2)}{f(\tilde r_{_{\rm{\uppercase\expandafter{\romannumeral1}}}})\tilde r_{_{\rm{\uppercase\expandafter{\romannumeral1}}}}}<0,\nonumber
\end{eqnarray}
and  meet the conditions, $\tilde h(\tilde r_{_{\rm{\uppercase\expandafter{\romannumeral1}}}})=f(\tilde r_{_{\rm{\uppercase\expandafter{\romannumeral1}}}})$, $\tilde h'(\tilde r_{_{\rm{\uppercase\expandafter{\romannumeral1}}}})=f'(\tilde r_{_{\rm{\uppercase\expandafter{\romannumeral1}}}})$, and
 $\tilde h''(\tilde r_{_{\rm{\uppercase\expandafter{\romannumeral1}}}})=f''(\tilde r_{_{\rm{\uppercase\expandafter{\romannumeral1}}}})$. From $T_{\rm H}(\tilde r_{_{\rm{\uppercase\expandafter{\romannumeral1}}}})>T_{\rm H}(r_{_{\rm{\uppercase\expandafter{\romannumeral2}}}})$, we deduce $\tilde\beta<\beta_2$. With this constraint and the assumption $\tilde\lambda>\lambda_2$ together we first consider the following four different situations which are drawn in Figure \ref{h_r_1} and Figure \ref{h_r_2}. %for the sketches of $h_1'(r)$ (the green curve) and $h_2'(r)$ (the red curve).

(i) In the range of parameters $\beta_1<\tilde\beta<\beta_2$ and $\lambda_2<\tilde\lambda<\lambda_1$, $\tilde h'(0)=\tilde \lambda b^2$ is between $h_1'(0)=\lambda_1 b^2$ and $h_2'(0)=\lambda_2 b^2$, so the point of tangency $\tilde r_{_{\rm{\uppercase\expandafter{\romannumeral1}}}}$, if it exists, should be between $r_{_{\rm{\uppercase\expandafter{\romannumeral1}}}}$ and $r_{_{\rm{\uppercase\expandafter{\romannumeral2}}}}$ in order to make the integral area surrounded by $\tilde h(\tilde r_{_{\rm{\uppercase\expandafter{\romannumeral1}}}})$ and $[0, \tilde r_{_{\rm{\uppercase\expandafter{\romannumeral1}}}}]$ be equal to the integral area surrounded by $f(\tilde r_{_{\rm{\uppercase\expandafter{\romannumeral1}}}})$ and $[0, \tilde r_{_{\rm{\uppercase\expandafter{\romannumeral1}}}}]$, rather than be larger than $r_{_{\rm{\uppercase\expandafter{\romannumeral2}}}}$ as supposed, see sketch (a) in Figure \ref{h_r_1}.

(ii) In the range of parameters $\beta_1<\tilde\beta<\beta_2$ and $\lambda_2<\lambda_1<\tilde\lambda$, the curve of $\tilde h'(r)$ lies above that of $h_1'(r)$, so there is no $\tilde r_{_{\rm{\uppercase\expandafter{\romannumeral1}}}}$ to satisfy the equality of integral areas about $\tilde h'(r)$ and $f'(r)$, see sketch (b) in Figure \ref{h_r_1}.

(iii) In the range of parameters $\tilde\beta<\beta_1<\beta_2$ and $\lambda_2<\tilde\lambda<\lambda_1$, the curve of $\tilde h'(r)$ lies below that of $h_1'(r)$, so we find no $\tilde r_{_{\rm{\uppercase\expandafter{\romannumeral1}}}}$ to satisfy the equality of integral areas about $\tilde h'(r)$ and $f'(r)$, see sketch (a) in Figure \ref{h_r_2}.

(iv) In the range of parameters $\tilde\beta<\beta_1<\beta_2$ and $\lambda_2<\lambda_1<\tilde\lambda$, since $\tilde h'(0)=\tilde\lambda$ is larger than $h_1'(0)=\lambda_1$, in order to ensure the equality of the integral areas, the curve of $\tilde h'(r)$ must intersect with $h_1'(r)$ before $h_1'(r)$ is tangent to $f'(r)$, which makes the point of tangency of $\tilde h'(r)$ and $f'(r)$ smaller than $r_{_{\rm{\uppercase\expandafter{\romannumeral1}}}}$ or even non-existent, see the sketch (b) in Figure \ref{h_r_2}.

In addition, for the situation $\tilde \lambda<\lambda_2$ where the curve of $\tilde h'(r)$ is below that of $h_2'(r)$, we find no points of tangency to meet $\tilde h(\tilde r_{_{\rm{\uppercase\expandafter{\romannumeral1}}}})=f(\tilde r_{_{\rm{\uppercase\expandafter{\romannumeral1}}}})$.

As analyzed in the above for both $\tilde \lambda>\lambda_2$ (four possible situations) and $\tilde \lambda<\lambda_2$, we can conclude that no other maximum of temperature exists. Moreover, no other minimum of temperature exists, either, since no other  minimum points are available after the minimum point $r_{_{\rm{\uppercase\expandafter{\romannumeral2}}}}$. Otherwise, the asymptotic behavior of the temperature at a large horizon radius would be violated. As a result, when $b$ is relatively large, there are only one maximum point and one minimum point on the curve of $T_{\rm H}$, which is similar to the cyan curves of Figures~\ref{fig:T_H_e} and \ref{T_H_s}.

Eq.~(\ref{T_H_b}) shows that $T_{\rm H}$ increases as $b$ decreases at fixed $r_+$. Eq.~(\ref{T'_G}) indicates that $T_{\rm H}'(r)$ is always positive for $r<r_1$, and that for $r>r_1$, it can also be positive for a small $b$. When the curve is considered as a whole to be varying smoothly with respect to $b$, there exist four types of situations: one single maximum, one maximum and one following minimum, one inflexion, and no extremal points on the curve. They appear in the order as the AdS radius $b$ decreases from infinity to zero,
% without considering the hoop conjecture,
which explains the similarity of thermodynamic properties in the models with a general mass distribution.
%However, for those general mass distributions with $b$ constrained by the hoop conjecture, such a similarity is not obvious and needs to be verified further.
%may still hold, see, for instance, Figures~\ref{fig:T_H_e} and \ref{T_H_s}.

\subsection{Entropy}
From eq.~(\ref{eq:dM}) and eq.~(\ref{eq:T_G}), the entropy at constant pressure is derived to be
\begin{align}
S=\int^{r_+}_{r_0}\frac{\mathrm{d}M}{T_H}=\int^{r_+}_{r_0}\frac{2\pi r}{f(r)}\mathrm{d}r,
\label{S_G}
\end{align}
revealing that the entropy depends on the mass distribution $f(r)$. Incidentally, it returns naturally to the ordinary Schwarzschild and Schwarzschild-AdS black holes if one sets $f(r)=1$.

When we make a further extension, starting from the non-singularity of metric eq.~(\ref{ds^2}) while neglecting the finiteness of density, the discussions above can be generalized to the case in which $f(r)$ is infinitesimal higher than first order at the origin. Specifically, when one takes $f(r)$ in the form,
\begin{align}
f(r)=1-\frac{l_0^2}{r^2+l_0^2},
\end{align}
the entropy turns out to be
\begin{align}
S=\pi (r_+^2-r_0^2)+2\pi l_0^2 \ln\left(\frac{r_+^2}{r_0^2}\right),
\end{align}
where the correction is logarithmic, as calculated in the holographic metric~\cite{1604.03263}. We thus include the result of ref.~\cite{1604.03263} as our special case.

\section{Summary}

In this paper, we calculate the Hawking temperature and the entropy of black holes, and analyze the phase transitions under different vacuum pressure for the self-regular Schwarzschild-AdS black hole with the hydrogen-atom-like and the collapsed shell mass distributions. In particular, we provide the fine phase structures that depend on the cosmological constant or the thermodynamic pressure. There are four types of phase structures that correspond to the zero, the relatively low, the critical, and the relatively high pressure, respectively, where the second type is further separated into three sub-types due to its complexity. Thanks to the existence of a minimal length, the models we studied are free of singularity, and their pathological outcome of  evaporation can be cured. Specifically, when the temperature is smaller than $T_{\rm{HP}}$, the black hole even with positive Gibbs free energy is unable to decay into the pure thermal radiation due to the existence of extreme configurations. In addition, the entropy satisfies the area law only at a large horizon radius but has an obvious deviation dependent on the mass distribution at the near-extremal scale. In order to give the reason that different mass-smeared schemes lead to similarities in thermodynamic properties~\cite{1410.1706}, we investigate a general model whose mass density is based on an analytic expression of the $\delta$-function and the mass distribution takes a kind of step functions with continuity. We find that all such models indeed have the similar properties thermodynamically.
%do not analyze Hawing-Page phase transition in general due to the dependence of vanishing Gibbs free energy on specific model, which remains to be further discussed in future's work. Other than that,

\section*{Competing Interests}

The authors declare that they have no competing interests regarding the publication of this paper.

\section*{Acknowledgments}
Y-GM would like to thank H. Nicolai of Max-Planck-Institut f\"ur Gravitationsphysik (Albert-Einstein-Institut) for kind hospitality. Y-MW would like to thank C. Liu for helpful discussions. This work was supported in part by the National Natural
Science Foundation of China under grant No.11675081.

\end{document}